# Quark distributions and gluon contents of eta and etaprime

Janardan P. Singh[1] and Aditya B. Patel
Physics Department, Faculty of Science, The M.S. University of Baroda, Vadodara - 390002, India
[1]email: janardanmsu@yahoo.com

**Abstract:** UsingQCD sum rule, we calculate valance quark distribution functions of $\eta$ and of $\eta'$ in chiral limit. Extrapolating the calculated quark distribution functions at both end regions, their first two moments have been calculated and compared with those of other hadrons.

## 1. Introduction

The quark-gluon structure of pseudoscalar mesons $\eta$ and $\eta'$ is of great interest both theoretically and phenomenologically [1,2]. Their structure and the mixing angle are characterized, among others, by SU(3)-flavor symmetry breaking, U(1)$_A$ anomaly of QCD, OZI-rule violation, etc. Their study can reveal some important nonperturbative aspects of QCD and its vacuum structure. To see the nontrivial flavor structure of the $\eta$-$\eta'$ mesons, we start by defining the matrix elements

$$\langle 0|J_{\mu 5}^a|P(p)\rangle = i f_P^a p_\mu \tag{1}$$

where the index $a = 8,0$ denotes the octet and singlet axial current respectively. In terms of u, d and s quark fields, the currents are defined by

$$J_{\mu 5}^8 = \tfrac{1}{\sqrt{6}}\left(\bar{u}\gamma_\mu\gamma_5 u + \bar{d}\gamma_\mu\gamma_5 d - 2\bar{s}\gamma_\mu\gamma_5 s\right) \tag{2a}$$

$$J_{\mu 5}^0 = \tfrac{1}{\sqrt{3}}\left(\bar{u}\gamma_\mu\gamma_5 u + \bar{d}\gamma_\mu\gamma_5 d + \bar{s}\gamma_\mu\gamma_5 s\right) \tag{2b}$$

The pseudoscalar meson $P$ can be $\eta$ or $\eta'$. Following current literature, the constants $f_P^a$ can be written in matrix form in terms of mixing angles $\theta_8$ and $\theta_0$.

$$\begin{pmatrix} f_\eta^8 & f_\eta^0 \\ f_{\eta'}^8 & f_{\eta'}^0 \end{pmatrix} = \begin{pmatrix} f_8 \cos\theta_8 & -f_0 \sin\theta_0 \\ f_8 \sin\theta_8 & f_0 \cos\theta_0 \end{pmatrix} \tag{3}$$

Since the matrix elements in Eq.(1) correspond to the annihilation of two quarks at one space-time point, the decay constants $f_P^a$ are related to the values of the light-cone wave functions 'at the origin'[1]. If we use the numerical values of $f_P^a$ determined, for example, in Ref.[3], then it gives enough hint of complex flavor structure of $\eta$ and $\eta'$. The anomaly in the divergence of singlet axial current is

$$\partial^\mu J_{\mu 5}^0 = \tfrac{2i}{\sqrt{3}}\left(m_u \bar{u}\gamma_\mu\gamma_5 u + m_d \bar{d}\gamma_\mu\gamma_5 d + m_s \bar{s}\gamma_\mu\gamma_5 s\right) - \tfrac{\sqrt{3}}{4}\tfrac{\alpha_s}{\pi} G\tilde{G} \tag{4}$$

where

$$G\tilde{G} = \frac{1}{2}\epsilon^{\mu\nu\rho\sigma}G_{\mu\nu}^a G_{\rho\sigma}^a, \quad \epsilon^{0123} = +1 \quad (5)$$

The anomaly in Eq. (4) induces a significant mixing between the $q\bar{q}$ and $s\bar{s}$ fields. For $m_{u,d} \ll m_s$, the matrix elements of anomaly between vacuum and pseudoscalar states are [4]

$$\left\langle 0 \left| \frac{3}{4}\frac{\alpha_s}{\pi} G\tilde{G} \right| \eta \right\rangle = -\sqrt{3}m_\eta^2 (f_8 cos\theta_8 - \sqrt{2}f_0 sin\theta_0) \quad (6a)$$

$$\left\langle 0 \left| \frac{3}{4}\frac{\alpha_s}{\pi} G\tilde{G} \right| \eta' \right\rangle = -\sqrt{3}m_{\eta'}^2 (f_8 sin\theta_8 + \sqrt{2}f_0 cos\theta_0) \quad (6b)$$

It is well known that in the vector meson sector the ω and φ are nearly pure $q\bar{q}$ and $s\bar{s}$ states, respectively. The smallness of φ-ω mixing angle is consistent with OZI-rule. On the other hand, in the pseudoscalar sector, the large mixing between the quark flavors can not be only due to quark-antiquark annihilation mechanism, and is interpreted as the result of nontrivial effect of the anomaly which is connected with the topological properties of the QCD vacuum. It has been suggested [5] that physical $\eta'$ arises as a result of mixing of SU(3) singlet $\eta_0$ with the topological field $Q\left(=\frac{\alpha_s}{8\pi}G\tilde{G}\right)$ and then the octet $\eta_8$.

Instanton is an essential feature of QCD and is often used to model its vacuum. It has been invoked to explain anomalously high masses of $\eta$ and $\eta'$ mesons and also for the remarkable differences observed in the mixing angle for the pseudoscalar mesons and vector mesons [6].

On phenomenological side, the quark-gluon structure of $\eta$ and $\eta'$ mesons is being seriously debated in current literature. Recent claim of relatively large glue content of $\eta'$ by KLOE collaboration [7] and by Kou [8] is in conflict with the analysis of radiative decays $V \to P\gamma$ and $P \to V\gamma$ involving $\eta$ and $\eta'$ by Escribano and Nodal [9] and Thomas [10] who find no evidence for gluonic contribution in $\eta$ or $\eta'$. As a consequence quark structure also differs in these analyses. Ke et.al. [11], using the data of semileptonic decays of D and $D_s$ have extracted the mixing angles of $\eta$-$\eta'$-$G$ system, but have not reached any definite conclusion on fraction of gluonium in $\eta$ and $\eta'$ because the experimental errors are relatively large.

In this paper, we study the quark-gluon structure of $\eta$ and $\eta'$ from the perspective of valance quark distribution functions. The experimental study of quark distribution function of mesons is limited to pions [12] and kaons [13], and the data is rather poor. Theoretical studies on meson structure functions have been done using QCD sum rule [14,15], Dyson-Schwinger equation [16], Nambu-Jona-Lasinio model [17], instanton model [18], light front constituent quark model [19], etc. In this work we will use QCD sum rule method which has been used for structure function calculation of baryons as well as mesons within a limited range of Bjorken variable [14,15,20].

The QCD sum rule has been used to calculate various hadronic characteristics in a model independent way with a good accuracy. Two-point correlation functions involving octet and singlet axial currents, pseudoscalar currents and axial anomaly has been successfully used to calculate decay constants and mixing angles of $\eta$ and $\eta'$[3], the derivative of topological susceptibility $\chi'(0)$, the mass of $\eta'$ in the chiral limit as well as singlet decay constant in the same limit [21]. Recently eta-nucleon coupling constant was determined by considering two-point correlation function between vacuum and one meson state [22]. All these determinations are in good accord with those determined by other methods. The QCD sum rule has also been used to determine the structure functions and quark distributions in photons and a few hadrons within a limited range of Bjorken variable. In this work we will use this method for determining quark distributions of $\eta$ and $\eta'$ and that, in turn, will be used to understand their quark-gluon structure. The method is based on consideration of 4-point correlator corresponding to forward scattering of two currents, one of which has the quantum numbers of the hadron of interest, and the other is electromagnetic or weak current. Since both $J_{\mu 5}^8$ and $J_{\mu 5}^0$ couple to $\eta$ and $\eta'$-state, we shall choose a combination of the two currents which couples either to $\eta$ or $\eta'$ in order to avoid interference.

## 2. Calculation method and analysis

To calculate the quark distributions in $\eta$ and $\eta'$ using QCD sum rule, we follow Ioffe and Oganesian[14,15]. First, to consider the case of $\eta$, we construct an axial current that couples to $\eta$ but not to $\eta'$:

$$J_\mu^{(\eta)} = \left(f_\eta^8 J_{\mu 5}^0 - f_\eta^0 J_{\mu 5}^8\right)/\left((f_\eta^8)^2 + (f_{\eta'}^0)^2\right)^{1/2} \qquad (7a)$$

$$= 0.593(\bar{u}\gamma_\mu\gamma_5 u + \bar{d}\gamma_\mu\gamma_5 d) - 0.544\bar{s}\gamma_\mu\gamma_5 s, \qquad (7b)$$

$$\langle 0|J_\mu^{(\eta)}|\eta'(p)\rangle = 0 \qquad (8a)$$

$$\langle 0|J_\mu^{(\eta)}|\eta(p)\rangle = if_\eta p_\mu \qquad (8b)$$

To get equation (7b), we have used numerical values for constants from Ref.[3]. Furthermore, using these numerical values we also evaluate $f_\eta \cong 161.3\ MeV$, which is not very different from $f_\eta \cong 169.5\ MeV$ [22] obtained for octet member $\eta_8$ from Gell-Mann-Okubo-like relations. Now consider the 4-point correlator [14] with two electromagnetic currents and two currents as given in Eqs.(7a, 7b):

$$\Pi_{\mu\nu\rho\sigma}(p_1, p_2; q_1, q_2) = \int e^{i(p_1 x + q_1 y - p_2 z)} d^4x d^4y d^4z$$
$$\times \langle 0|T\{J_\lambda^{(\eta)}(x) J_\mu^{em}(y) J_\nu^{em}(0) J_\rho^{(\eta)}(z)\}|0\rangle \qquad (9)$$

Among the various tensor structures of $\Pi_{\mu\nu\rho\sigma}$, the structure $(p_\mu p_\nu p_\lambda p_\rho/p \cdot q) \cdot \tilde{\Pi}(p^2, q^2, x)$ (on choosing $p_1 = p_2 = p$) is the most suitable for the study of the structure function. Here $x = -q^2/2p \cdot q$ is the Bjorken variable and the imaginary part $Im\tilde{\Pi}(p^2, q^2, x)$ in the s-channel is related to the $\eta$ structure function $F_{2\eta}(x)$.

For $p_1^2 \neq p_2^2$, the dispersion representation has the form

$$Im\ \tilde{\Pi}(p_1^2, p_2^2, x) = a(x) + \int_0^\infty \frac{\varphi(x,u)}{(u-p_1^2)} du + \int_0^\infty \frac{\varphi(x,u)}{(u-p_2^2)} du + \int_0^\infty du_1 \int_0^\infty du_2 \frac{\rho(x,u_1,u_2)}{(u_1-p_1^2)(u_2-p_2^2)} \qquad (10)$$

Borel transform is defined by

$$\mathcal{B}_{M^2} f(q^2) = \lim_{\substack{-q^2 \to \infty, \\ n \to \infty,\ -\frac{q^2}{n}=M^2}} \frac{(-q^2)^{n+1}}{n!} \left(\frac{d}{dq^2}\right)^n f(q^2) \qquad (11)$$

Applying a double Borel transformation to Eq.(10) in $p_1^2$ and $p_2^2$ variable removes the first three terms in the r.h.s of Eq.(10) giving

$$\mathfrak{B}_{M_1^2}\mathfrak{B}_{M_2^2} Im\widetilde{\Pi}(p_1^2, p_2^2, x) = \int_0^\infty du_1 \int_0^\infty du_2\, \rho(x, u_1, u_2) \exp\left[-\frac{u_1}{M_1^2} - \frac{u_2}{M_2^2}\right] \qquad (12)$$

It is convenient to put $M_1^2 = M_2^2 = 2M^2$[23]. In the region $u_1 < s_0$ and $u_2 < s_0$, where $s_0$ is the continuum threshold to be chosen for each channel, the spectral function has the form

$$\rho(x, u_1, u_2) = 2\pi F_{2\eta}(x) f_\eta^2 \delta(u_1 - m_\eta^2)\delta(u_2 - m_\eta^2) \qquad (13)$$

On the other hand for the region $u_1, u_2 > s_0$, the non-perturbative contributions may be neglected and the spectral function is described by quark loop spectral function $\rho^0(u_1, u_2, x)$. The other two regions when one of $u_1$ and $u_2$ is larger than $s_0$ while the other is smaller than $s_0$ correspond to the transition $\eta \to$ continuum. It is clear from Eq.(12) that for these three regions, there is an exponential suppression. Using the hypothesis of quark-hadron duality, the contributions of these regions is estimated to be contribution of the bare quark loop in the same regions, and the value of these contributions will be demanded to be less than approximately one-third of the entire contribution. $Im\widetilde{\Pi}(p_1^2, p_2^2, x)$ can also be calculated in QCD. This leads to the following equations,

$$\mathfrak{B}_{M_1^2}\mathfrak{B}_{M_2^2} Im\Pi_{QCD}^0 + Power\ correction = 2\pi F_{2\eta}(x) f_\eta^2 \exp[-m_\eta^2/M^2] \qquad (14)$$

$$\mathfrak{B}_{M_1^2}\mathfrak{B}_{M_2^2} Im\Pi_{QCD}^0 = \int_0^{s_0} du_1 \int_0^{s_0} du_2\, \rho^0(u_1, u_2, x) \exp\left[-\frac{(u_1+u_2)}{2M^2}\right] \qquad (15)$$

where $\rho^0(u_1, u_2, x)$ is the bare loop spectral function. The sum rule for u-quark distribution in η can be written as[14]:

$$xu_\eta(x) = 0.352 \frac{3}{2\pi^2} \frac{M^2}{f_\eta^2} x^2(1-x)e^{m_\eta^2/M^2} \left[\left(1 + \left(\frac{\alpha_s(M^2)\ln(Q^2/M^2)}{3\pi}\right)\right)\left(\frac{1+4x\ln(1-x)}{x} - \frac{2(1-2x)\ln(1-x)}{(1-x)}\right)\right)\left(1 - e^{\frac{-s_0}{M^2}}\right) - \frac{4\pi\alpha_s(Q^2)\cdot 4\pi\alpha_s(M^2) a^2}{(2\pi)^4 \cdot 3^7 \cdot 2^6 \cdot M^6} \frac{\omega(x)}{x^3(1-x)^3}\right] \qquad (16)$$

where the first numerical factor in r.h.s of Eq.(16) comes from the square of the numerical factor of the u-quark axial current in Eq.(7), $a = -(2\pi)^2\langle\bar{u}u\rangle$, $Q^2 = -q^2$ is the point where the sum rule is calculated and

$$\omega(x) = (5784x^4 - 1140x^3 - 20196x^2 + 20628x - 8292)ln2$$
$$+ 4740x^4 + 8847x^3 + 2066x^2 - 2553x + 1416 \qquad (17)$$

The first term in the square bracket on the r.h.s. in Eq.(16) is the perturbative contribution which has been checked by us to be correct. The expression containing $e^{\frac{-s_0}{M^2}}$ in the perturbative term is the contribution from non-resonance region on the phenomenological side parameterized as the contribution of the bare quark loop in the region $u_1, u_2 > s_0$ and transferred on the QCD side. The second term in the square bracket in the Eq.(16) is the nonperturbative contribution of dimension six with a very large number of Feynman diagrams and has been taken from Ref.[14].

Evidently, we have $d_\eta(x) = u_\eta(x)$. For s-quark distribution function, in addition to the perturbative and nonperturbative contributions appearing in Eq.(16), we also include the contribution due to finite mass of s-quark but retaining contribution up to $O(m_s^2)$ in the perturbative term:

$$xs_\eta(x) = \frac{0.296\times 3}{2\pi^2}\frac{M^2 e^{m_\eta^2/M^2}}{f_\eta^2}\left\{x^2(1-x)\left[\left(1+\left(\frac{\alpha_s(M^2)\ln(Q^2/M^2)}{3\pi}\right)\left(\frac{1+4x\ln(1-x)}{x}-\frac{2(1-2x)\ln(1-x)}{(1-x)}\right)\right)\left(1-e^{\frac{-s_0}{M^2}}\right) - \frac{4\pi\alpha_s(Q^2)\cdot 4\pi\alpha_s(M^2)(0.8)^2 a^2}{(2\pi)^4\cdot 3^7\cdot 2^6\cdot M^6}\frac{\omega(x)}{x^3(1-x)^3}\right] - \frac{xm_s^2}{M^2}\right\} \quad (18)$$

where $\langle\bar{s}s\rangle \cong 0.8\langle\bar{u}u\rangle$ has been used. The quark condensate of strange quarks has been found to be somewhat different from light quarks: $\langle\bar{s}s\rangle/\langle\bar{u}u\rangle$ =0.8±0.1 [24]. Unlike the case of pion, we have kept $\eta$-mass nonzero in evaluation of sum rule Eqs.(16) and (18). It has been observed [14] that above expressions for $u_\eta(x)$ and $s_\eta(x)$ are valid for $0.15 \leq x \leq 0.7$. In the standard QCD sum rule approach, the upper limit on $M^2$ is chosen so that the contribution of the continuum region is not more than 50% and the lower limit of $M^2$ is chosen so that the nonperturbative contribution of highest dimensional operator is not more than 10%. With our choice of $M^2$ and $s_0$, as given below, both these conditions are well satisfied.

The quark distribution functions $u_\eta(x)$ and $s_\eta(x)$ can be extrapolated in the region, where Eqs.(16) and (18) are inapplicable as follows: $u_\eta(x) \sim x^{-1/2}$ for $x \leq 0.15$ according to Regge behavior for low $x$ and $u_\eta(x) \sim (1-x)^2$ for $x \geq 0.7$ according to the quark counting rules for $x$ close to 1. However, since with these choices the extrapolations do not smoothly join the derived expressions for $xu_\eta(x)$ and $xs_\eta(x)$, we make the following modifications in extrapolations: $xu_\eta(x) \sim x^{1/2}(1-x)^\beta$, $xs_\eta(x) \sim x^{1/2}(1-x)^\gamma$ for $x \leq 0.15$ and $xu_\eta(x) \sim x^\alpha(1-x)^2$, $xs_\eta(x) \sim x^\delta(1-x)^2$ for $x \geq 0.7$, and choose the unknown exponents such that they smoothly join the respective curves at the boundaries. A general parameterization of the type $C_a x^{\sigma_a}(1-x)^{\mu_a}$ for the structure function $F_a(x, Q^2)$ had been used long back by Feynman and Field [25]. In recent years, the initial parton distributions for proton have been used which are of the form $xf_i(x, Q_0^2) = A_i x^{-\alpha_i}(1-x)^{\beta_i}(1+\gamma_i\sqrt{x}+\delta_i x)$ for the entire range of $x$ [26]. We find that $\beta$ =−5.25, $\gamma$ =−5.57, while $\alpha$ =3.80 and $\delta$ =3.56 make the matching of the curves with the extrapolations smooth. We have chosen $\alpha_s(1\,GeV^2) = 0.43$ and using two-loop formula we obtain $\alpha_s(2\,GeV^2)$ =0.33 with $\Lambda_{QCD}$= 0.315 GeV. We also use $m_s$=0.13 GeV and $\alpha_s(1\,GeV^2)a^2 = 0.13\,GeV^6$ [14]. Further, we have chosen $M^2$ =1 GeV$^2$ and $s_0 = 1.2\,GeV^2$ for which continuum contribution remains less than approximately 30% of the entire contribution. The contribution of the last term in Eq.(16), the nonperturbative contribution, is a few percent of the whole. In figures.(1a) and (1b), we have shown the behavior of $xu_\eta(x)$ and $xs_\eta(x)$ as a function of $M^2$ for $x = 0.15, 0.3, 0.5, 0.7$ and Q$^2$=2 GeV$^2$. Observing that these functions are approximately stable for variation against $M^2$ within the region $0.8\,GeV^2 < M^2 < 1.5\,GeV^2$, we calculate the first two moments

$Q_1^{(\eta)} = \int_0^1 q_\eta(x)dx$
and
$Q_2^{(\eta)} = \int_0^1 xq_\eta(x)dx$

for $q = u, s$; we have also done this by varying the boundaries of applicability of the region in $x$ as $x_1 = 0.15 \pm 0.03$ and $x_2 = 0.7 \pm 0.05$ and displayed the results in table 1. We have shown $xu_\eta(x)$ and $xs_\eta(x)$ (as given by Eq.(16) and Eq.(18)) in figure 2 and these functions along with their extrapolations in figure 3. We find that Evidently, we have $d_\eta(x) = u_\eta(x)$. For s-quark distribution function, in addition to the perturbative and nonperturbative contributions appearing in Eq.(16), we also include the contribution due to finite mass of s-quark but retaining contribution up to $O(m_s^2)$ in the perturbative term:

$$xs_\eta(x) = \frac{0.296\times 3}{2\pi^2}\frac{M^2 e^{m_\eta^2/M^2}}{f_\eta^2}\left\{x^2(1-x)\left[\left(1+\left(\frac{\alpha_s(M^2)\ln(Q^2/M^2)}{3\pi}\right)\left(\frac{1+4x\ln(1-x)}{x}-\frac{2(1-2x)\ln(1-x)}{(1-x)}\right)\right)\left(1-e^{\frac{-s_0}{M^2}}\right) - \frac{4\pi\alpha_s(Q^2)\cdot 4\pi\alpha_s(M^2)(0.8)^2 a^2}{(2\pi)^4\cdot 3^7\cdot 2^6\cdot M^6}\frac{\omega(x)}{x^3(1-x)^3}\right] - \frac{xm_s^2}{M^2}\right\} \quad (18)$$

where $\langle\bar{s}s\rangle \cong 0.8\langle\bar{u}u\rangle$ has been used. The quark condensate of strange quarks has been found to be somewhat different from light quarks: $\langle\bar{s}s\rangle/\langle\bar{u}u\rangle$ =0.8±0.1 [24]. Unlike the case of pion, we have kept $\eta$-mass nonzero in evaluation of sum rule Eqs.(16) and (18). It has been observed [14] that above expressions for $u_\eta(x)$ and $s_\eta(x)$ are valid for $0.15 \leq x \leq 0.7$. In the standard QCD sum rule approach, the upper limit on $M^2$ is chosen so that the contribution of the continuum region is not more than 50% and the lower limit of $M^2$ is chosen so that the nonperturbative contribution of highest dimensional operator is not more than 10%. With our choice of $M^2$ and $s_0$, as given below, both these conditions are well satisfied.

The quark distribution functions $u_\eta(x)$ and $s_\eta(x)$ can be extrapolated in the region, where Eqs.(16) and (18) are inapplicable as follows: $u_\eta(x) \sim x^{-1/2}$ for $x \leq 0.15$ according to Regge behavior for low $x$ and $u_\eta(x) \sim (1-x)^2$ for $x \geq 0.7$ according to the quark counting rules for $x$ close to 1. However, since with these choices the extrapolations do not smoothly join the derived expressions for $xu_\eta(x)$ and $xs_\eta(x)$, we make the following modifications in extrapolations: $xu_\eta(x) \sim x^{1/2}(1-x)^\beta$, $xs_\eta(x) \sim x^{1/2}(1-x)^\gamma$ for $x \leq 0.15$ and $xu_\eta(x) \sim x^\alpha(1-x)^2$, $xs_\eta(x) \sim x^\delta(1-x)^2$ for $x \geq 0.7$, and choose the unknown exponents such that they smoothly join the respective curves at the boundaries. A general parameterization of the type $C_a x^{\sigma_a}(1-x)^{\mu_a}$ for the structure function $F_a(x, Q^2)$ had been used long back by Feynman and Field [25]. In recent years, the initial parton distributions for proton have been used which are of the form $xf_i(x, Q_0^2) = A_i x^{-\alpha_i}(1-x)^{\beta_i}(1+\gamma_i\sqrt{x}+\delta_i x)$ for the entire range of $x$ [26]. We find that $\beta$ =−5.25, $\gamma$ =−5.57, while $\alpha$ =3.80 and $\delta$ =3.56 make the matching of the curves with the extrapolations smooth. We have chosen $\alpha_s(1\,GeV^2) = 0.43$ and using two-loop formula we obtain $\alpha_s(2\,GeV^2)$ =0.33 with $\Lambda_{QCD}$= 0.315 GeV. We also use $m_s$=0.13 GeV and $\alpha_s(1\,GeV^2)a^2 = 0.13\,GeV^6$ [14]. Further, we have chosen M$^2$ =1 GeV$^2$ and $s_0 = 1.2\,GeV^2$ for which continuum contribution remains less than approximately 30% of the entire contribution. The contribution of the last term in Eq.(16), the nonperturbative contribution, is a few percent of the whole. In figures.(1a) and (1b), we have shown the behavior of $xu_\eta(x)$ and $xs_\eta(x)$ as a function of $M^2$ for $x = 0.15, 0.3, 0.5, 0.7$ and Q$^2$=2 GeV$^2$. Observing that these functions are approximately stable for variation against $M^2$ within the region $0.8\,GeV^2 < M^2 < 1.5\,GeV^2$, we calculate the first two moments

$Q_1^{(\eta)} = \int_0^1 q_\eta(x)dx$
and
$Q_2^{(\eta)} = \int_0^1 xq_\eta(x)dx$

for $q = u, s$; we have also done this by varying the boundaries of applicability of the region in $x$ as $x_1 = 0.15 \pm 0.03$ and $x_2 = 0.7 \pm 0.05$ and displayed the results in table 1. We have shown $xu_\eta(x)$ and $xs_\eta(x)$ (as given by Eq.(16) and Eq.(18)) in figure 2 and these functions along with their extrapolations in figure 3. We find that

$$M_1^{(\eta)} \equiv \int_0^1 [u_\eta(x) + d_\eta(x) + s_\eta(x)]dx = 0.95 \pm 0.02 \tag{19}$$

$$M_2^{(\eta)} \equiv \int_0^1 [u_\eta(x) + d_\eta(x) + s_\eta(x)]xdx = 0.39 \pm 0.01 \tag{20}$$

where the uncertainties in the numerical values arise due to variations in the boundaries of the applicability of the region in $x$ as stated above. The corresponding central result for pion has been 0.84 and 0.21 respectively [14].

**Table 1**: First and second moments of quark distribution functions in $\eta$ at $Q^2 = 2.0\ GeV^2$ with the extrapolating functions incorporated; $x_1$ and $x_2$ are the boundaries beyond which extrapolations have been used. Notations used in the first column are defined in the text.

| Moments of quark distr. funct. | $x_1 = 0.12$ $x_2 = 0.65$ | $x_1 = 0.12$ $x_2 = 0.70$ | $x_1 = 0.12$ $x_2 = 0.75$ | $x_1 = 0.15$ $x_2 = 0.65$ | $x_1 = 0.15$ $x_2 = 0.70$ | $x_1 = 0.15$ $x_2 = 0.75$ | $x_1 = 0.18$ $x_2 = 0.65$ | $x_1 = 0.18$ $x_2 = 0.70$ | $x_1 = 0.18$ $x_2 = 0.75$ |
|---|---|---|---|---|---|---|---|---|---|
| $U_1^{(\eta)}$ | 0.339 | 0.342 | 0.344 | 0.342 | 0.344 | 0.346 | 0.348 | 0.350 | 0.352 |
| $S_1^{(\eta)}$ | 0.258 | 0.252 | 0.253 | 0.252 | 0.253 | 0.254 | 0.256 | 0.258 | 0.258 |
| $M_1^{(\eta)}$ | 0.929 | 0.936 | 0.941 | 0.935 | 0.941 | 0.946 | 0.951 | 0.958 | 0.962 |
| $U_2^{(\eta)}$ | 0.140 | 0.143 | 0.144 | 0.141 | 0.143 | 0.144 | 0.141 | 0.143 | 0.145 |
| $S_2^{(\eta)}$ | 0.106 | 0.105 | 0.106 | 0.122 | 0.105 | 0.106 | 0.104 | 0.105 | 0.106 |
| $M_2^{(\eta)}$ | 0.385 | 0.390 | 0.394 | 0.403 | 0.391 | 0.395 | 0.386 | 0.391 | 0.395 |

We observe that the changes in the values of $x_1 = 0.15 \pm 0.03$ and $x_2 = 0.7 \pm 0.05$ results in changes of $M_1^{(\eta)}$ and $M_2^{(\eta)}$ which are a couple of percent while the contribution to $M_2^{(\eta)}$ from the extrapolated region is estimated to be $\cong 20\%$ of the whole. We also note that the numerical value of $M_1^{(\eta)}$ is close to unity in spite of extrapolations giving credence to the correctness of this approach. We expect error in the value of $M_2^{(\eta)}$ due to the extrapolations to be maximum 10%. As stated earlier, contribution due to the resonance states in the model used is about 30% of the whole. We guess error due to the modeling also to be at most 10%. Nonperturbative contribution appearing as the last term in Eq. (16) is just a few percent. We expect the contribution due to the leftover nonperturbative and perturbative terms in expressions (16) and (18), and that due to the uncertainties in values of constants $\alpha_s$, $m_s$, $f_\eta$ as well as due to the choice of the Borel parameter $M^2$ and the continuum threshold $s_0$ to give an error up to 5%. Thus, we expect the error in our result for $M_2^{(\eta)}$ to be at most 25% which is a typical error one encounters in QCD sum rule calculations [3,22]:

$$M_2^{(\eta)} = 0.39 \pm 0.1 \tag{21}$$

QCD evolution equation for the moments to leading order are given by [27]

$$\frac{M_2^{(\eta)}(Q^2)}{M_2^{(\eta)}(\mu^2)} = \left\{\frac{\alpha_s(Q^2)}{\alpha_s(\mu^2)}\right\}^{d_2} \tag{22}$$

with the anomalous dimension $d_2 = -2A_2/b_0$, $A_2 = -16/9$ and $b_0 = 11N_C/3 - 2n_f/3$ while for $M_1^{(\eta)}$ there is no evolution. In evolving to higher energies, we make the matching between $n_f = 5$ and 4 at $\mu = 5$ GeV and between $n_f = 4$ and 3 at $\mu = 1.777$ GeV [28]. We have tabulated $M_2^{(\eta)}(Q^2)$ for some typical energy scales obtained from evolution of the central value written in Eq. (21) in table 2.

**Table2:** Evolution of $M_2^{(\eta)}(Q^2)$ with $Q^2$ for some typical energy values.

| $Q^2$ (GeV$^2$) | 2 | 25 | 100 | 200 | $M_Z^2$ |
|---|---|---|---|---|---|
| $M_2^{(\eta)}(Q^2)$ | 0.39 | 0.33 | 0.30 | 0.29 | 0.24 |

We can proceed to discuss the quark distribution functions for $\eta'$ in a way similar to what we did for $\eta$. We construct a quark current $J_\mu^{(\eta')}$ which couples to $\eta'$ but not to $\eta$:

$$\langle 0 | J_\mu^{(\eta')} | \eta(p) \rangle = 0 \tag{23}$$

$$\langle 0 | J_\mu^{(\eta')} | \eta'(p) \rangle = i f_{\eta'} p_\mu \tag{24}$$

$$J_\mu^{(\eta')} = (f_\eta^8 J_{\mu 5}^0 - f_\eta^0 J_{\mu 5}^8)/((f_\eta^8)^2 + (f_\eta^0)^2)^{1/2} \tag{25a}$$

$$= 0.499(\bar{u}\gamma_\mu\gamma_5 u + \bar{d}\gamma_\mu\gamma_5 d) + 0.709 \bar{s}\gamma_\mu\gamma_5 s \tag{25b}$$

$$f_{\eta'} = 154.25\ MeV \tag{25c}$$

Following the same procedure which was followed for $\eta$, we find that the sum of the integrated probabilities of the u, d and s quarks $M_1^{(\eta')} \cong 1.94$ while the total momentum carried by the valence quarks and antiquarks as a fraction of $\eta'$ momentum $2M_2^{(\eta')} > 1$. Here, clearly the method fails. We can see from Eq. (16) as to how this happens. The difference for the case of $\eta'$ comes from $m_{\eta'}^2$ appearing in the exponential and $f_{\eta'}^2$ appearing in the denominator with the result that both contribute for increment of the numerical value of $u_{\eta'}(x)$ and $s_{\eta'}(x)$. The major reason for this unphysical behavior of the quark probability distribution function is clearly the anomalously large value of $\eta'$ mass, which is $U(1)_A$ problem of QCD. There is nothing in the present approach which compensates for this increment and hence this physically unacceptable result.

There have been several suggestions as to how the $\eta'$ acquires its unexpectedly large mass. According to Kogut and Susskind a contribution to the $\eta'$ mass in the chiral limit (this would have been zero for a Goldstone boson) could be obtained by the mixing between two infrared enhanced gluons, with momentum space propagator $D(k) \sim 1/k^4$ for $k^2 \to 0$ [29]. 't Hooft had shown that instantons lead to the non-conservation of the axial charge and so induce a $2N_f$ fermion operator which gives rise to a non-zero $\eta'$-mass in the chiral limit [30]. Witten and Veneziano have proposed that in the large $N_c$ limit, where $N_c$ is the number of colors, the anomalous mass of $\eta'$ is related to the topological susceptibility of the gluodynamics [31, 32]. In recent years solutions to the Dyson-Schwinger equation for the propagators of QCD and the quark-gluon vertex have been employed with the Kogut-Susskind mechanism to determine the anomalous mass of the $\eta'$ is the chiral limit [33]. In QCD sum rule approach the mass of $\eta'$ in the chiral limit and its decay constants in the same limit were determined from the study of anomaly-anomaly correlator [21].

The problem in QCD sum rule approach based on two point correlator involving singlet axial current has been noticed earlier [34,3]. A solution to this invoking instantons was suggested in Ref.[3], but that was of ad hoc nature and it will not be used here. Since $\eta'$ is largely singlet, the appearance of the problem with $\eta'$ is not unexpected. Nevertheless, we will calculate $q_{\eta'}(x)$ in the chiral limit for which we use [21]

$$m_{\eta'} \cong 723 MeV, f_{\eta'} \cong 178\ MeV \qquad (26)$$

Since the octet of pseudoscalars has become massless in chiral limit, there is no mixing. We have repeated the analysis of Eq. (16) for mass and coupling as given in Eq. (26). Behavior of $xq_{\eta'}(x)$ as a function of $M^2$, for few select values of $x$, has been shown in figure4. In figure 5, we have shown the calculated $xq_{\eta'}(x)$ as a function of $x$, and in figure 6, we have shown the calculated $xq_{\eta'}(x)$ along with its extrapolations. From this, we get sensible results for quark number density $q_{\eta'}(x)$ whose first and second moments are shown in table 3. Similarity of the quark distribution function in this case with that for $\eta$ meson is consistent with the universality of the $q\bar{q}$ wave functions of $\pi^0, \eta$ and $\eta'$ [35]. The results obtained for the moments

$$M_1^{(\eta')} = 1.03 \pm 0.06, \qquad M_2^{(\eta')} = 0.42 \pm 0.01 \qquad (27)$$

are now sensible and close to those obtained for $\eta$. Thus, for $\eta'$ meson also around 80% of momentum is carried by the quarks while remaining $\approx 20\%$ of its momentum is carried by gluons albeit in the chiral limit. We expect $s_{\eta'}(x)$ to decrease for non-zero $m_s$ as was observed for the case of $\eta$ and this decrement should be larger in a realistic case since the share of s-quark current is larger in $J_\mu^{(\eta')}$ (see Eq. (25b)). This is also supported by the fact that the numerical value of the matrix element $\left\langle 0 \left| a_\eta \frac{\alpha_s}{\pi} G\tilde{G} \right| \eta' \right\rangle$ is known to get increased roughly by 1.5 times for the physical states compared to the corresponding value in the chiral limit [36]. The contribution of the extrapolated part of $xq_{\eta'}(x)$ curve has a larger share $\approx 29\%$ in this case. Coupled with the fact that in this case our analysis has been done in the chiral limit, the uncertainty in the result for $M_2^{(\eta')}$ for a physical $\eta'$ will be larger – could be as high as 35%. Clearly the evolution of $M_2^{(\eta')}$, calculated here in the chiral limit, would be similar to that of $M_2^{(\eta)}$.

**Table 3**: First and second moments of quark distribution function at $Q^2 = 2.0\ GeV^2$ in $\eta'$ with the extrapolating functions incorporated; $x_1$ and $x_2$ are the boundaries beyond which extrapolations have been used. Notations are as in table 1.

| Moments of quark distr. funct. | $x_1 = 0.12$ $x_2 = 0.65$ | $x_1 = 0.12$ $x_2 = 0.70$ | $x_1 = 0.12$ $x_2 = 0.75$ | $x_1 = 0.15$ $x_2 = 0.65$ | $x_1 = 0.15$ $x_2 = 0.70$ | $x_1 = 0.15$ $x_2 = 0.75$ | $x_1 = 0.18$ $x_2 = 0.65$ | $x_1 = 0.18$ $x_2 = 0.70$ | $x_1 = 0.18$ $x_2 = 0.75$ |
|---|---|---|---|---|---|---|---|---|---|
| $U_1^{(\eta')}$ | 0.327 | 0.330 | 0.328 | 0.323 | 0.326 | 0.326 | 0.338 | 0.341 | 0.363 |
| $M_1^{(\eta')}$ | 0.981 | 0.988 | 0.983 | 0.969 | 0.977 | 0.983 | 1.015 | 1.023 | 1.090 |
| $U_2^{(\eta')}$ | 0.137 | 0.139 | 0.137 | 0.136 | 0.138 | 0.140 | 0.137 | 0.139 | 0.142 |
| $M_2^{(\eta')}$ | 0.410 | 0.416 | 0.411 | 0.409 | 0.415 | 0.420 | 0.411 | 0.417 | 0.426 |

### 3. Discussion and conclusion

We have obtained a nontrivial result that the sum of the first moments of the all the valance quark distributions, $M_1 \approx 1$ for the $\eta$ meson and for the $\eta'$ meson in the chiral limit. On the other hand, the sum of the second moments of all the valance quark and antiquark distributions, $2M_2 \approx 0.8$ for both the cases. According to the conventional logic it means that about 80% of the momentum of the meson (in chiral limit for $\eta'$) is carried by all the valance quarks and antiquarks whereas the remaining 20% of the meson's momentum is carried by the gluons which are electrically neutral. We also expect that this latter figure may increase for $\eta'$ when finite quark masses are taken. In this sense, the cases of $\eta$ and $\eta'$ are more like longitudinally polarized $\rho$ meson for which case also only about 20% of the momentum has been found to be carried by gluons and sea quarks [15]. It may be remarked that charged pions and transverse $\rho$ mesons behave like a nucleon where about 50% of the total momentum is carried by gluons and sea quarks at moderately high energies [15]. On the other hand, $\eta$- $\eta'$ system may contain valance gluons created by topological charge density operator $Q$. This is evident from the fact that the expectation value of this operator between vacuum and one $\eta$-or $\eta'$-state is non-zero (see Eqs.(6a, 6b)). This corresponds to annihilation of two, three or four gluons at one space-time point and matrix elements are related to the values of the corresponding light-cone wave functions 'at the origin'. A measure of the strength of gluonic coupling to quark coupling for $\eta$ and $\eta'$ states can be given as follows: Neglecting $m_u$ and $m_d$ compared to $m_s$, we have for the $\eta$-current

$$\partial^\mu J_\mu^{(\eta)} = -1.088\ i\ m_s \bar{s}\gamma_5 s - 0.080 \frac{\alpha_s}{\pi} G\tilde{G} \qquad (28)$$

using the constants from Ref.[3]. The ratio of the matrix elements of the gluonic part to the quark part in the $\eta$ state:

$$0.08 \left\langle 0\left|\frac{\alpha_s}{\pi} G\tilde{G}\right|\eta\right\rangle / (1.088\ i\ m_s \langle 0|\bar{s}\gamma_5 s|\eta\rangle) \cong 0.32 \qquad (29)$$

On the other hand, for the $\eta'$-current, we have

$$\partial^\mu J_\mu^{(\eta')} = 1.418\ i\ m_s \bar{s}\gamma_5 s - 0.213 \frac{\alpha_s}{\pi} G\tilde{G} \qquad (30)$$

and the ratio of the matrix elements of the gluonic part to the quark part in the $\eta'$-state is

$$-0.213 \left\langle 0 \left| \frac{\alpha_s}{\pi} G\tilde{G} \right| \eta' \right\rangle / (1.418\, i\, m_s \langle 0|\bar{s}\gamma_5 s|\eta'\rangle) \cong 1.33 \qquad (31)$$

Comparison of Eqs. (29) and (31) indicates qualitatively a larger gluonic content of $\eta'$. However, we can not give a more quantitative estimate of this. It may also be noted that there are low energy effective chiral Lagrangians [37,38] in which SU(3) flavor singlet pseudoscalar $\eta_0$ is coupled to topological charge density $Q$. These Lagrangians have been used to study low energy processes involving the η and $\eta'$ [39]. Phenomenological evidence for the gluon content of these mesons have also been given using decay processes involving them [40]. A sizable gluonium content of $\eta'$ could help to understand the unexpected high value of the branching ratio for $B \to \eta' K_s$ decay [41]. A particularly interesting investigation from the point of view of inelastic scattering is that of two-gluon Fock components occurring in hard processes involving the $\eta$ and $\eta'$ mesons to leading twist accuracy [42]. Thus, we see that the η-$\eta'$ system is unique among the well established hadrons on account of having substantial coupling to a purely gluonic operator.

It has been shown that instantons provide qualitative resolution of the large mass of the $\eta'$ meson [43], and it may be argued that possibly they contribute a large fraction of the momentum carried by $\eta'$. However, estimates by Baulieu et al. [44] show that the instanton correction for deep inelastic scattering are utterly negligible for $Q^2 \gtrsim 1$ GeV$^2$.

η and $\eta'$ mesons are unique among the pseudoscalars for they have anomalously large masses and it is more so for $\eta'$. The large masses of these mesons are believed to be contributed by gluons. Gluons play important role in several processes involving production and decay of η and $\eta'$. Large matrix elements of gluonic operators between vacuum and one $\eta'$-state indicate presence of significant valence gluonic component in $\eta'$. However, when it comes to the momentum fraction carried by gluons in fast moving mesons, it is ≈20% at $Q^2 \simeq 2$ GeV$^2$ for both η and $\eta'$ just like a longitudinal ρ meson, and it is at very high energies $\sim M_Z^2$ that this fraction evolves to 50%. In conclusion, QCD sum rule provides a viable theoretical approach for calculation of quark distribution functions of η and $\eta'$ mesons. We believe our determination of parton distribution functions of these mesons can be useful as initial inputs for further studies of these functions.

**Acknowledgement**: The authors gratefully acknowledge the financial assistance from DST, New Delhi.

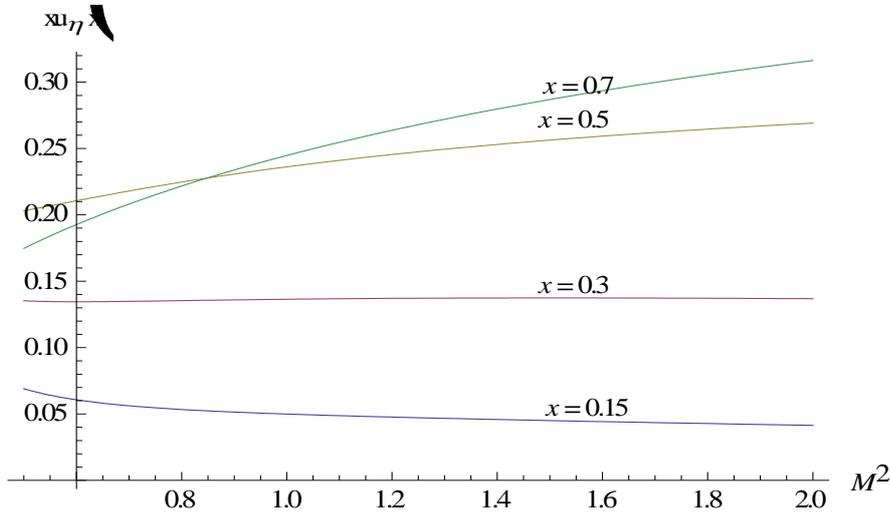

Fig. 1a: u-quark distribution function for fixed values of x at $Q^2=2$ GeV$^2$ as a function of Borel mass parameter for $\eta$ meson. The largest variation is for the curve with x=0.7 which falls from 0.287 for $M^2= 1.5$ GeV$^2$ to 0.222 for $M^2= 0.8$ GeV$^2$.

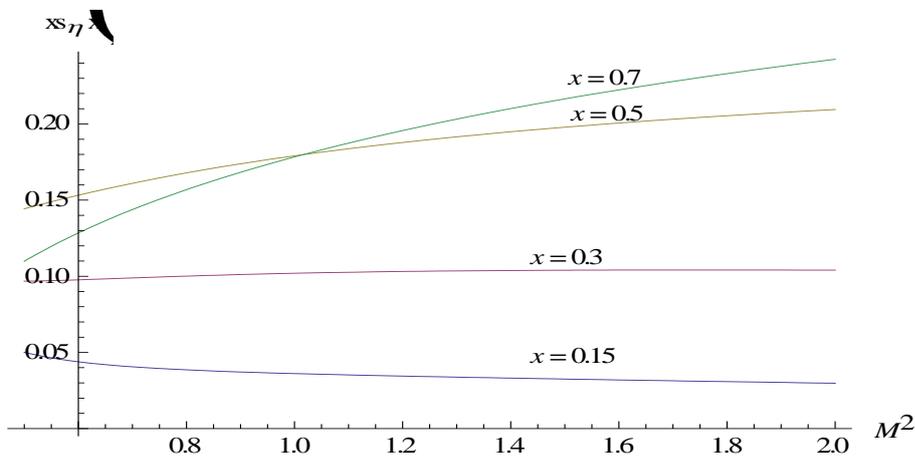

Fig. 1b: s-quark distribution function for fixed values of x at $Q^2=2$ GeV$^2$ as a function of Borel mass parameter for $\eta$ meson. The largest variation is for the curve with x=0.7 which falls from 0.216 for $M^2= 1.5$ GeV$^2$ to 0.157 for $M^2= 0.8$ GeV$^2$.

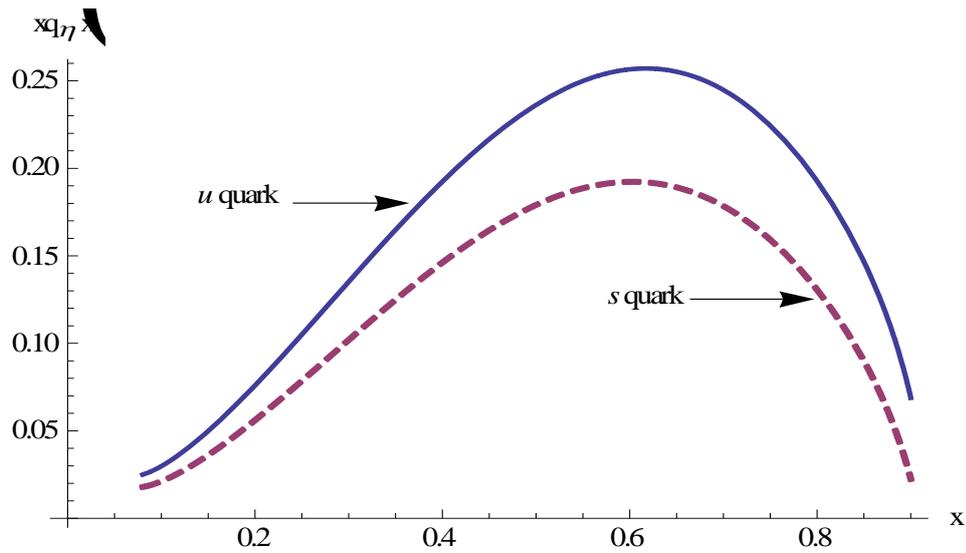

Fig.2: u-quark and s-quark distribution functions, as given by Eq.(16) and Eq.(18) respectively, at $Q^2=2$ GeV$^2$ for $\eta$ meson.

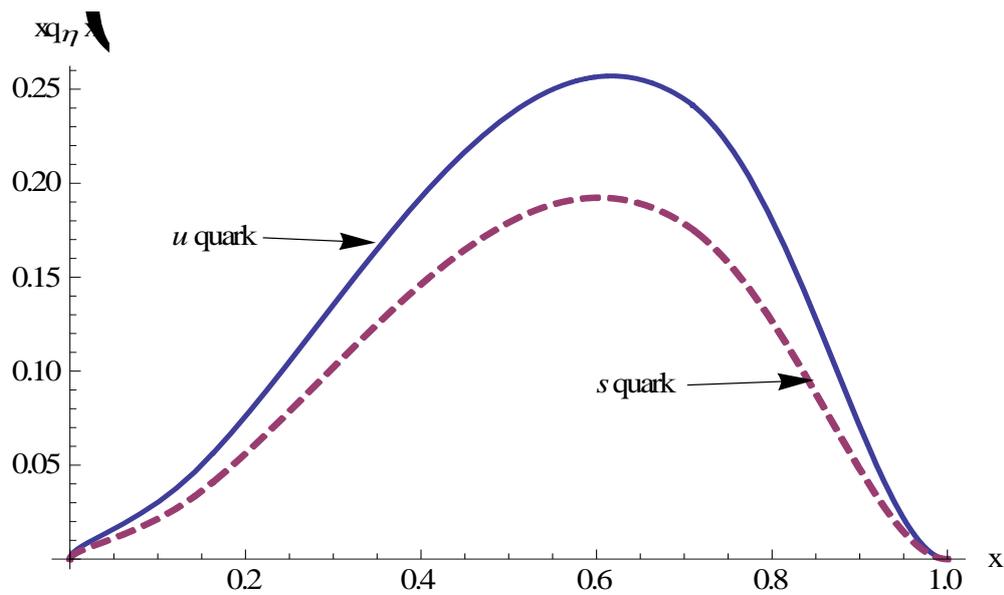

Fig.3: u-quark and s-quark distribution functions along with their extrapolations, as given in the text, in the region $x < 0.15$ and $x > 0.7$ at $Q^2=2$ GeV$^2$ for $\eta$ meson.

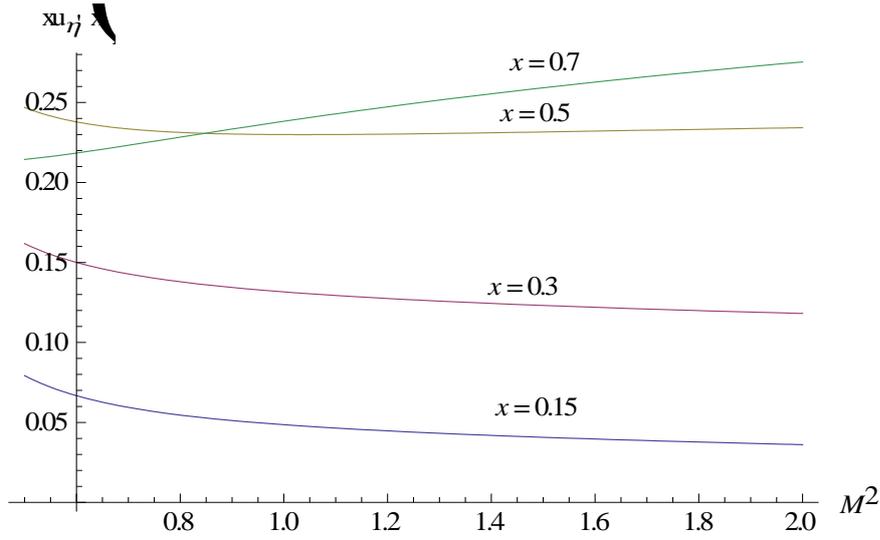

Fig.4: Quark distribution function for fixed values of x at $Q^2=2$ GeV$^2$ as a function of Borel mass parameter for $\eta'$ meson. The largest variation is for the curve with x=0.7 which falls from 0.259 for $M^2=1.5$ GeV$^2$ to 0.228 for $M^2=0.8$ GeV$^2$.

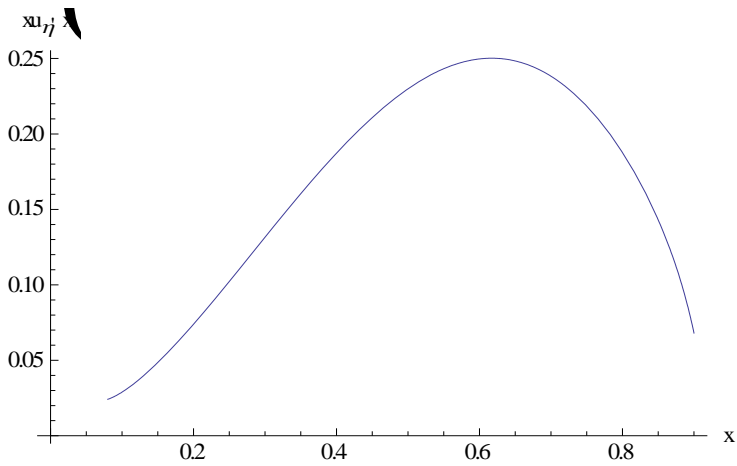

Fig. 5: Quark distribution function as given by Eq.(16), but with mass and coupling given in Eq.(26), at $Q^2=2$ GeV$^2$ for $\eta'$ meson and overall numerical factor 1/3 in place of 0.352.

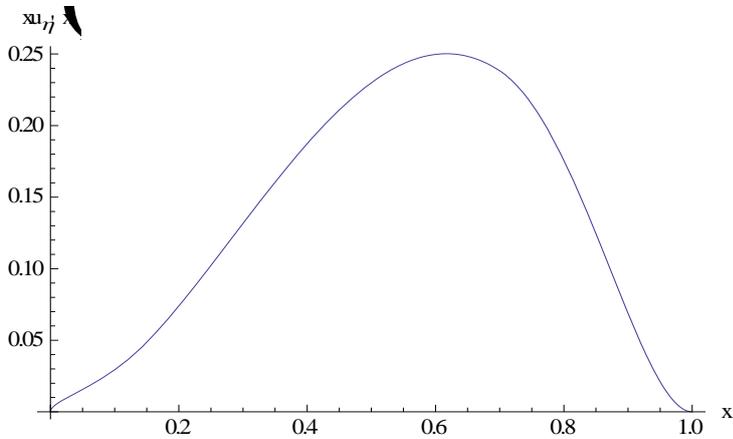

Fig. 6: Quark distribution function along with its extrapolations, as given in the text, in the region $x < 0.15$ and $x > 0.7$ at $Q^2=2$GeV$^2$ for $\eta'$ meson.

**References**

1. Feldman T 2000 *Int. J. Mod. Phys.* A **15**, 159
2. Shore G M 2006 *Nucl. Phys.* B **744**, 34
3. Singh J P and Pasupathy J 2009 *Phys. Rev*. D **79**, 116005
4. Akhoury R and Frere J M 1989 *Phys. Lett* B **220**, 258
5. Shore G M and Veneziano G 1988 *Nucl. Phys*. B **516**, 333
6. 'tHooft G 1999 *Preprint* hep-th/9903189.
7. Ambrosino F et al 2007 *Phys. Lett*. B **648**, 267
   Ambrosino F et al 2009 J. High Energy Phys. *JHEP* **0907** 105
8. Kou E 2001 *Phys. Rev*. D **63**, 054027
9. Escribano R and Nadal J 2007 J. High Energy Phys. *JHEP* **0705**, 006
10. Thomas C E 2007 J. High Energy Phys. *JHEP* **071**0, 026
11. Ke H-W, Yuan X-H and Li X-Q 2011 arXiv: 1101.3407.
12. Betev B et al [NA10 Collaboration] 1985 *Z. Phys. C***28**, 15
    Conway J S et al 1989 *Phys. Rev*. D **39**, 92
    Badier J et al. [NA3 Collaboration] 1983 *Z. Phys. C* **18**, 281
13. Badier J et al 1980 *Phys. Lett*. B **93**, 354
14. Ioffe B L and Oganesian A G 2000 *Eur. Phys. J. C* **13**, 485
15. Ioffe B L and Oganesian A G 2001 *Phys. Rev*. D **63**, 096006
16. Hecht M B et al 2001 *Phys. Rev*. C **63**, 025213
17. Shigetani T, Suzuki K and Toki H 1993 *Phys. Lett*. B **308**, 383
18. Dorokhov A E and Tomio L 2000*Phys. Rev*. D **62**, 014013
19. Fredrico T and Miller G A 1994 *Phys. Rev*. D **50**, 201
20. Belyaev V M and Ioffe B L 1988 Nucl. Phys. B **310**, 548
    Ioffe B L and Oganesian A G 2003 Nucl.Phys. A **714**, 145
21. Pasupathy J, Singh J P, Singh R K and Upadhyay A 2006 *Phys. Lett*. B **634**, 508
22. Singh J P, Lee F X and WangL 2011 *Int. J. Mod. Phys.* A **26**, 947
23. Ioffe B L and Smilga AV 1983 Nucl. Phys. B **216**, 373
24. Ioffe B L 1981 Nucl. Phys. B **188**, 317; Errata 1982 Nucl.Phys. B **191**, 591.
25. Feynman R P and Field R D 1977 Phys. Rev. D **15**, 2590
26. Martin A D, Stirling W J, Thorne E S and Watt G 2007 Phys. Lett. B **652**, 292
27. Close F E, DonnachieS and Shaw G 2007 Electromagnetic Interactions and Hadronic Structure, Cambridge Univ. Press
28. Yndurain F J 2006 The Theory of Quark and Gluon Interactions, Springer
29. Kogut J P and Susskind L 1974 *Phys. Rev*. D **10** 3468
30. 'tHooft G 1976 *Phys. Rev. Lett.* **37**, 8
    'tHooft G 1986 *Phys. Rep.* **142**, 357
31. Witten E 1979 *Nucl. Phys*. B **156**, 269
32. Veneziano G 1979 *Nucl. Phys.* B **159**, 213
    Veneziano G 1980 *Phys. Lett*. B **95**, 90
33. Alkofer R, Fischer C S and Williams R 2008 Eur. Phys. J. A **38**, 53
34. Ioffe B L andSamsonov A V 2000 Yad. Fiz. **63**, 1527
35. Anisovich V V, Melikhov D I and Nikonov V A 1997 Phys. Rev. D **55**, 2918
36. Cheng H-Y, Li H and Liu K F 2009 Phys. Rev. D **79**, 014024
37. Rosenzweig C, Schechter J and Trahern C G 1980 *Phys. Rev*. D **21**, 3388
    Nath P and Arnowitt R 1981 *Phys. Rev*. D **23**, 473
38. Vecchia P Di and Veneziano G 1980 *Nucl. Phys.* B **171**, 253
39. Bass S D 1999 *Phys. Lett*. B **463**, 286
    Bass D S 2009 *Acta Phys. Polon. Supp.* **2**, 11
40. Ball P, Frere J M and Tytgat M 1996 *Phys. Lett*. B **365**, 367
41. Charng Y-Y, Kurimoto T and Li H 2006 *Phys. Rev*. D **74**, 074024
42. Kroll P and Passek-Kumericki K 2003 *Phys. Rev*. D **67**, 054017
    Ali A and Parkhomenko A Ya 2003 Eur. Phys. J. C **30**, 367; 2004 Eur. Phys. J C **33**, S518
    Agaev S S and Stefanis N G 2004 Phys. Rev. D **70**, 054020
43. Gogohia V 2001 Phys. Lett. B **501**, 60
44. Baulieu L, Ellis J, Gaillard M K and Zaknewski W J 1979 Phys. Lett. B **81**, 224